\documentclass[pdftex]{article}

\usepackage{arxiv}

\RequirePackage{amsmath}

\usepackage[utf8]{inputenc} % allow utf-8 input
\usepackage[T1]{fontenc}    % use 8-bit T1 fonts
\usepackage{hyperref}       % hyperlinks
\usepackage{url}            % simple URL typesetting
\usepackage{booktabs}       % professional-quality tables
\usepackage{amsfonts}       % blackboard math symbols
\usepackage{nicefrac}       % compact symbols for 1/2, etc.
\usepackage{microtype}      % microtypography
\usepackage{cleveref}       % smart cross-referencing
\usepackage{lipsum}         % Can be removed after putting your text content
\usepackage{natbib}
\usepackage{doi}

\RequirePackage{fancybox}
\RequirePackage[dvips]{graphicx}
\RequirePackage{amssymb}

\RequirePackage{float}

\RequirePackage[shortlabels]{enumitem}

\RequirePackage{comment}

\RequirePackage{algpseudocode,algorithm}%,algorithmicx}
\RequirePackage{array}

\crefname{figure}{Fig.}{Figs.}
\Crefname{figure}{Figure}{Figures}
\crefname{table}{Tab.}{Tabs.}
\Crefname{table}{Tab.}{Tabs.}
\crefname{equation}{Eq.}{Eqs.}
\Crefname{equation}{Eq.}{Eqs.}
\crefname{section}{Sec.}{Secs.}
\Crefname{section}{Sec.}{Secs.}
\crefname{subsection}{Sec.}{Secs.}
\Crefname{subsection}{Sec.}{Secs.}
\crefname{subsubsection}{Sec.}{Secs.}
\Crefname{subsubsection}{Sec.}{Secs.}
\usepackage[misc]{ifsym}
\usepackage{soul}
\usepackage[normalem]{ulem}

\DeclareGraphicsExtensions{ps}
\makeatletter
  \newcommand\figcaption{\def\@captype{figure}\caption}
  \newcommand\tabcaption{\def\@captype{table}\caption}
\makeatother
\def\mi{\begin{equation}}
\def\mf{\end{equation}}

\def\mia#1\mfa{\begin{align}#1\end{align}}%se puede poner sin eqn number align*
\def\miar#1\mfar{\begin{eqnarray}#1\end{eqnarray}}
\def\mmi#1\mmf{\begin{multline}#1\end{multline}}
\def\<#1\>{\begin{equation}#1\end{equation}}
\def\[#1\]{\begin{equation}#1\end{equation}} % this overwrites the original setting without eqn number
\def\reff#1{(\ref{#1})}

\renewcommand{\v}[1]{\ensuremath{\mathbf{#1}}} % for vectors
\newcommand{\gv}[1]{\ensuremath{\mbox{\boldmath$ #1 $}}}  % for vectors of Greek letters
\let\baraccent=\= % rename builtin command \= to \baraccent
\renewcommand{\=}[1]{\stackrel{#1}{=}} % for putting numbers above =

\newcommand*{\plot}[2][]{\noindent\includegraphics[width=\ifx &#1& 0.98\else #1\fi\linewidth]{#2}\vskip -0.8cm\figcaption{}}

\RequirePackage[normalem]{ulem}

\newtheorem{remark}{Remark}
\usepackage{tikz}
\usetikzlibrary{arrows.meta, arrows, bending}
\title{Data-Driven Crowd Dynamics using Kinetic Theory and Ensemble-based Data Assimilation}

\author{ \hspace{1mm}Santiago ~Rosa\\
 	Facultad de Matemática, Astronomía, Física y Computación, Universidad Nacional de Córdoba\\
 	Instituto de Física Enrique Gaviola, CONICET\\
 	santiago.rosa@mi.unc.edu.ar
    \AND
    \hspace{1mm}Manuel ~Pulido\\
 	Facultad de Ciencias Exactas, Naturales y Agrimensura, Universidad Nacional del Nordeste\\
 	Instituto de Modelado e Innovación Tecnológica, CONICET\\
 	\AND
    \hspace{1mm}Orlando ~Billoni\\
 	Facultad de Matemática, Astronomía, Física y Computación, Universidad Nacional de Córdoba\\
 	Instituto de Física Enrique Gaviola, CONICET\\
 	\AND
    \hspace{1mm}Juan Martín ~Guerrieri\\
 	Facultad de Ciencias Exactas, Naturales y Agrimensura, Universidad Nacional del Nordeste\\
 	Instituto de Modelado e Innovación Tecnológica, CONICET\\
 	\AND
    \hspace{1mm}Juan Pablo ~Agnelli\\
 	Facultad de Matemática, Astronomía, Física y Computación, Universidad Nacional de Córdoba\\
 	Centro de Investigación y Estudios de Matemática, CONICET
}

\hypersetup{
pdftitle={Data-Driven Crowd Dynamics using Kinetic Theory and Ensemble-based Data Assimilation},
pdfauthor={Santiago ~Rosa, Manuel ~Pulido, Orlando ~Billoni, Juan M. ~Guerrieri, Juan P. ~Agnelli},
pdfkeywords={Data assimilation, Kinetic theory},
}

\begin{document}
\graphicspath{{./figuras/}}

\maketitle

\begin{abstract}
	Understanding pedestrian dynamics is critical for mitigating crowd-related risks and improving public safety. In this work, we propose a data-driven mesoscopic modeling framework that combines the kinetic theory of active particles with data assimilation techniques. The framework uses an ensemble Kalman filter to sequentially estimate the time-dependent spatial distribution of pedestrians and model parameters by fusing observations with the mesoscopic forward model state. Through a series of twin experiments, we show that the panic factor---a key behavioral parameter---is identifiable within this framework. We also evaluate the robustness of the approach by assimilating synthetic observations generated by an agent-based model (ABM). This setup introduces structural model error because the ABM is governed by microscopic rules that differ fundamentally from the mesoscopic kinetic equations. Despite this discrepancy, the ensemble Kalman filter, through the observation-based innovation term, successfully drives the kinetic model to track the observed pedestrian density while simultaneously recovering the panic factor. In this framework, observations act as a physical constraint on the evolution of the kinetic model. Crowd-dynamics models, including ABMs and kinetic models, often rely on phenomenological terms to describe social interactions, with parameters that are highly uncertain. Our findings indicate that in such systems, free-running simulations inevitably may diverge from the true state, whereas an online data-driven approach effectively constrains the system's trajectory to its underlying dynamical manifold.
\end{abstract}

% keywords can be removed

\section{Introduction}
In recent years, the modeling of crowd dynamics has emerged as an essential tool for urban planning and public safety \cite{Zhang23,Jiang25}. It is used to quantify the complex emergent behaviors that arise when individual decision-making interacts with spatial and social constraints, especially during high-stress situations such as fires or accidents. During panic events, social interactions between individuals are altered and can be represented as a function of psychological stress or fear, leading to competitive behavior that produces abrupt slowdowns or flow jamming \cite{Helbing00,Helbing05}. Models can predict the formation of dangerous crowding at exit points, allowing engineers to design structural interventions that facilitate more efficient evacuation during disasters \cite{Shiwacoti13}.

Pedestrian dynamics can be modeled from a microscopic to a macroscopic scale perspective \cite{Hughes02}. Agent-Based Models (ABMs) simulate crowds as systems of autonomous agents with distinct characteristics and behavioral rules, such as the social force model \cite{Helbing95} or cellular automata \cite{Li19}. In contrast, macroscopic hydrodynamic models treat a crowd as a fluid, defining state variables such as density and velocity fields whose the evolution is governed by the continuity equation \cite{Hughes02}. These models are particularly effective at identifying shock waves \cite{Johansson08}, which are localized surges in density that propagate through a crowd and are observed during disasters.

%,  providing a view of how high-pressure zones are produced across the domain..

In addition,  mesoscopic models can be regarded as intermediate between the microscopic and macroscopic scales. In particular, the kinetic theory for active particles (KTAP) \cite{Arlotti2002} provides a mathematical framework for describing the evolution of complex systems of interacting individuals, and has been successfully applied to crowd dynamics \cite{BellomoBellouquid11,Bellomo13,Agnelli14,KIM19}. 
A key feature distinguishing KTAP 
from classical kinetic theories for inert matter is that the interacting entities, referred to as \emph{active particles}, possess an internal micro-state that evolves over time.
This evolution is governed by stochastic, game-theoretic interaction rules designed to capture behavioral traits, rather than purely mechanical interactions.
Such a formulation enables the modeling of complex phenomena typical of living systems, including social contagion (e.g., the spread of fear or stress~\cite{Kim21,Agnelli25}).
Consequently, KTAP provides a powerful framework for the analysis of high-density evacuation scenarios, where human behavior and irrational responses may play a crucial role~\cite{Agnelli14}.

To bridge the gap between theoretical simulations and real-world observations, recent studies have increasingly adopted data-driven approaches to model pedestrian dynamics. 
Pouw et al. \cite{Pouw24} utilized a model based on Langevin dynamics to simulate pedestrian movements across various scenarios, calibrating it with empirical data via statistical field theories to parameterize interaction potentials. More recently,
Kim et al. (2025) \cite{Kim25} considered a kinetic model of crowd evacuation 
proposed in Agnelli et al. (2014) \cite{Agnelli14}, and formulated an inverse problem to estimate a panic parameter that plays a key role in the model. This approach can be viewed as a constrained optimization problem, in which the kinetic model acts as a constraint, and the objective functional measures the mismatch between the observed  and the model-predicted densities. 

Ensemble-based data assimilation (DA) offers a statistical framework for synchronizing simulations with sequential observations, enabling continuous state updates while accounting for uncertainty \cite{carrassi2018data}. In particular, the Ensemble Kalman Filter (EnKF) \cite{burguers98} enables parameter estimation via state augmentation \cite{Ruiz13}. 
By treating uncertain model parameters as part of the evolving state vector, the system can dynamically self-calibrate the parameters considering their covariances with the state evolution.
While prior work has applied DA to ABMs of pedestrian dynamics \cite{Ward16,clay21,Cocucci22}, the application of the data assimilation framework to mesoscopic kinetic models remains largely underexplored.

This work proposes a data-driven technique based on the Localized Ensemble Transform Kalman Filter (LETKF \cite{hunt07}), which combines pedestrian trajectory observations such as those obtained from video surveillance, with the mesoscopic model of Agnelli et al. (2014) \cite{Agnelli14}, to jointly estimate the crowd density field and the model's panic factor. Similar to Kim et al. (2025) \cite{Kim25}, we estimate a key behavioral parameter for a  mesoscopic kinetic model. However, our approach differs in several aspects. First, since we employ ensemble-based data assimilation with an augmented state, this enables the simultaneous real-time estimation of the system state and the model parameters. %so that the state is continously synchronized  toward the observed pedestrian dynamics as new data arrive. 
Second, the ensemble formulation explicitly accounts for uncertainties in the observations, model predictions and parameters. In addition, our method does not require a known reference solution, remaining effective in the presence of noisy observational data.  Moreover, it can operate successfully with a surrogated model under the presence of structural model errors\cite{Ruiz15}, while the system state is continuously synchronized with the observed pedestrian dynamics. This makes the framework particularly well suited for realistic scenarios, where data are inherently noisy and the underlying pedestrian dynamics are only partially captured by the model. 

The outline of this work is as follows. \Cref{metodos} describes the kinetic model and introduces the data assimilation framework.  \Cref{detalles_exp} details the numerical solver and the  experimental setup, including a twin experiment and an experiment in  which synthetic observations are generated from an ABM simulation. \Cref{resultados} presents and discusses the results of both experiments. Finally, in \Cref{conclusiones} we draw the conclusions.

%%%------------------------------------------------------------------------------------------------------------------------------------------------------------
\section{Methods} \label{metodos}
%%--------------------------------------------------------------------------
%Model

%----------------------------------------------------------------
\subsection{A spatial kinetic model of crowd evacuation} \label{modelo}
%-----------------------------------------------------------------

This work focuses on data-driven constraints applied to a KTAP framework that models crowd evacuation dynamics. We adopt the model introduced by Agnelli et al. (2014)~\cite{Agnelli14}, which describes the evacuation of a crowd from a bounded domain with specified exits. Within this model, pedestrians are treated as active particles whose micro-states are defined by their position and velocity direction. The global state of the system is characterized by a distribution function over these micro-states. Furthermore, the model accounts for velocity transitions ---affecting both speed and direction---governed by a specific decision-making strategy.

%----------------------------------------------------------------
%\subsubsection{Model notation}
%----------------------------------------------------------------

\begin{figure}[htbp]
    \centering
    \resizebox{0.85\textwidth}{!}{
        \begin{tikzpicture}[scale=0.45]
        \useasboundingbox (-2,-2) rectangle (31,23);
        %%%% DOMAIN %%%%
        % Lines
        \draw (1,0) -- (28,0);      % bottom
        \draw (1,20) -- (28,20);    % top
        \draw (0,1) -- (0,19);      % left
        \draw (29,19) -- (29,13);   % right upper
        \draw (29,7.5) -- (29,1);   % right lower
        % Rounded corners (approximation)
        \draw (0,1) arc[start angle=180,end angle=270,radius=1];
        \draw (1,20) arc[start angle=90,end angle=180,radius=1];
        \draw (29,1) arc[start angle=0,end angle=-90,radius=1];
        \draw (29,19) arc[start angle=0,end angle=90,radius=1];
        % Labels        
        \node at (2,1.5) { \mbox{\large $\Omega$}};
        \node at (-1,-1) {\mbox{\large$\partial \Omega$}};
        \node at (30,-1) {\mbox{\large$W$}};
        \node at (30,10) {\mbox{\large$E$}};
        %%%% VECTORS %%%%		
        % Exit vector
        \fill (16,3) circle (0.5);
        \draw[-{Latex},very thick] (16,3) -- (22,5.1);
        \draw[dashed] (16,3) -- (29,7.5);
        \node at (18,5.5) {$\vec{\nu}(\mathbf{r})$};
        \node at (14.5,3) {$\mathbf{r}$};
        \node at (26,4) {$d_E(\mathbf{r})$};
        % Tangent vector
        \fill (9,12) circle (0.5);
        \draw[-{Latex},very thick] (9,12) -- (13,16);
        \node at (7.9,12.3) {$\mathbf{\bar{r}}$};
        \draw[dashed] (9,12) -- (15,12);
        \draw (11.5,12) arc[start angle=0, end angle=45, radius=2.5]; % arc
        \node at (13,13.35) {$\theta_\ell$};
        \draw[dashed] (13,16) -- (17,20);   % continuation
        \node at (9.5,17.8) {$d_W(\mathbf{\bar{r}},\theta_\ell)$};
        \fill (17,20) circle (0.5);        % collision point
        \node at (15.3,21.5) {$\mathbf{\bar{r}}_W$};
        \draw[-{Latex},very thick] (17,20) -- (23,20);
        \node at (20.5,21.5) {${\vec{\tau}}(\mathbf{\bar{r}},\theta_\ell)$};	
        \end{tikzpicture}
    }
    \caption{
	    Geometry of the bounded domain $\Omega$ with boundary $\partial \Omega = W \cup E$. 
        For a pedestrian located at $\mathbf{r}$, $d_E(\mathbf{r})$ denotes the shortest distance to the exit, and $\vec{\nu}(\mathbf{r})$ is the unit vector pointing toward it. A pedestrian at $\mathbf{\bar{r}}$ moving in direction $\theta_\ell$ is expected to collide with the wall at $\mathbf{\bar{r}}_W$; it then selects the tangential direction, $\vec{\tau}(\mathbf{\bar{r}},\theta_\ell)$, to $\partial\Omega$ at $\mathbf{\bar{r}}_W$ that leads toward the exit.
            }
    \label{fig:1-domain}
\end{figure}
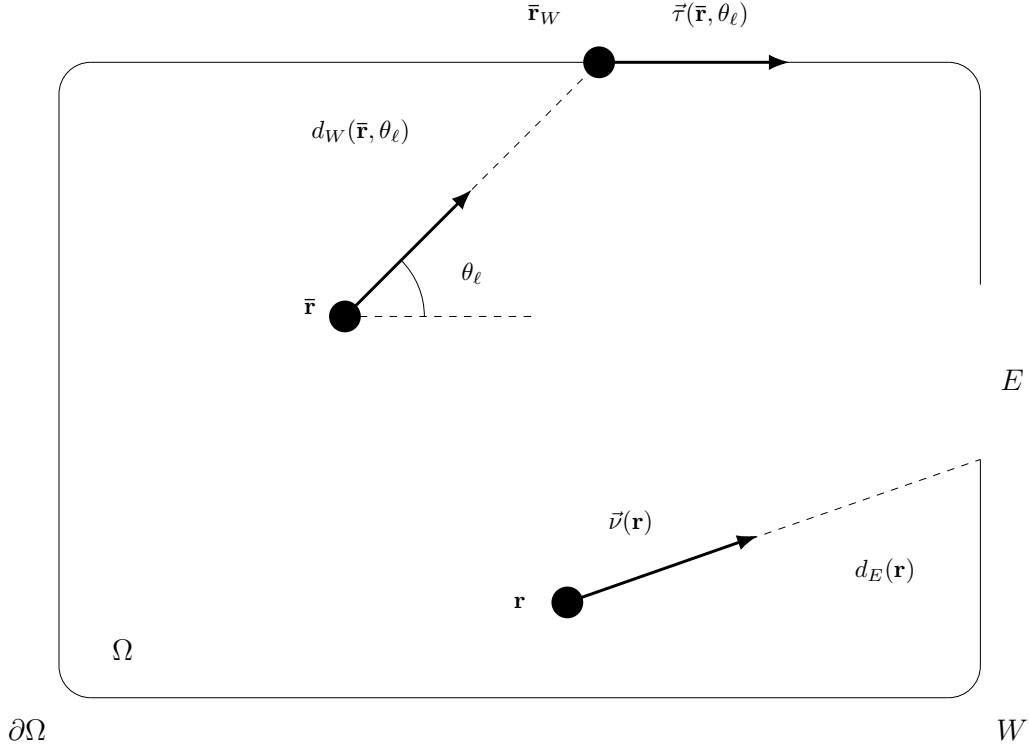

Following Agnelli et al. (2014)~\cite{Agnelli14}, we consider pedestrian motion within a bounded spatial domain, denoted by $\Omega \subset \mathbb{R}^2$. The boundary of this domain, $\partial \Omega$, is partitioned into a prescribed exit region $E \subset \partial \Omega$ and the remaining portion, which constitutes the wall, $W \subset \partial \Omega$, as illustrated in \cref{fig:1-domain}.

Each pedestrian is considered  an active particle, having an intrinsic microscopic state.
Consistent with the framework established in Agnelli et al. (2014)~\cite{Agnelli14}, the model incorporates hybrid continuous-discrete features. Specifically, the position vector 
$\mathbf{r} = (r_x,r_y)$ is treated as a continuous variable defined in $\Omega$. 
The velocity is expressed in polar coordinates as $\mathbf{v} = v~(\cos \theta,\sin \theta)$. The velocity magnitude, $v$, is modeled as a continuous, deterministic variable whose evolution is governed by macroscopic influences derived from the overall system dynamics.
Conversely, the direction $\theta$ is modeled as a discrete variable that can assume values within a defined finite set
\begin{equation}
\label{Ith}
I_{\theta}= \left\{\theta_h =\frac{h-1}{N_\theta} 2\pi\,:\,h=1, \ldots, N_\theta \right\},
\end{equation}
where $N_\theta$ denotes its cardinality.

The overall state of the pedestrians is characterized by the distribution function
$f_h(t,\mathbf{r})=f(t,\mathbf{r},\theta_h)$.
Under suitable integrability conditions, $f_h(t,\mathbf{r})d\mathbf{r}$ represents the number of individuals who, at time $t$  and possessing velocity  direction $\theta_h$, are located within the infinitesimal rectangle $[r_x, r_x + d r_x] \times [r_y, r_y + dr_y]$.
From this distribution function, various macroscopic quantities of interest can be derived. For instance, the local pedestrian density at position $\mathbf{r}$ is obtained by
\begin{equation}\label{eq:rho} 
\rho(t,\mathbf{r}) = \sum_{h=1}^{N_\theta} f_{h}(t,\mathbf{r}), 
\end{equation}
while the total population within the entire domain is given by
\begin{equation} 
N(t) = \int_\Omega \rho(t,\mathbf{r}) d\mathbf{r}.
\label{eq:Npop_norm}
\end{equation}

All physical variables  are non-dimensionalized by scaling them with suitable characteristic scales. In particular, let $L$ denote the domain diameter, $V_M$ the maximum pedestrian speed under optimal conditions, $T=L/V_M$ the reference time, and $\rho_M$ the maximum admissible density. The corresponding dimensionless variables are defined as
$\hat{\v r}=\v r/L$, $\hat{t}=t/T$, $\hat{v}=v/V_M$, and $\hat{f}=f/\rho_M$, although the hats are omitted hereafter for simplicity. Under this normalization, the local density defined in \cref{eq:rho} is bounded by 1, representing the maximal attainable density.

%%%%%%%%%%% Figure 1 %%%%%%%%%%%%%%%%%%%%%%%%%%%%%%%%%
The proposed framework accounts for both pedestrian--environment and pedestrian--pedestrian interactions. The underlying mathematical structure is derived from a mass balance of particles within an elementary volume of the  phase space. 
By considering the net flux into this volume, arising from both transport phenomena and interaction dynamics, we obtain the  following system of partial differential equations  
\begin{equation}\label{eq:general-structure}
\partial_{t}f_h(t,\mathbf{r}) + \mathrm{div}_\mathbf{r}(\mathbf{v}_h[\rho](t,\mathbf{r})f_h(t,\mathbf{r}) ) =  \mathcal{J}_h [f](t,\mathbf{r}),  
\end{equation}
for $h=1,\dots,N_\theta$, with $\mathbf{v}_h[\rho] = v[\rho](\cos\theta_h,\sin\theta_h)$. Here, the square brackets denote functional dependence on the local density $\rho$ or on its spatial gradient.
The left-hand side describes the transport of pedestrians, while the source term $\mathcal{J}_h[f]$ quantifies the net change in the distribution of pedestrians moving in direction $\theta_h$ due to behavioral interactions.

The model proposes that pedestrians adjust their speed based on local congestion levels; specifically,  the velocity magnitude $v$ is a function of the local density $\rho$. Consequently, higher densities reduce walking speeds, while lower densities allow for faster pedestrian movement. In particular, pedestrians move at a maximal speed 
within low-density regimes (free-flow conditions) up to a critical density threshold $\rho_c$. For $\rho > \rho_c$, the speed decreases progressively to zero, which is modeled through a polynomial dependence on $\rho$; see Sec.~2.3 of Agnelli et al. (2014)~\cite{Agnelli14}.

Regarding velocity directions, pedestrian movement is modeled as a non-deterministic decision-making process.
Rather than following rigid rules, pedestrians update their walking 
directions by participating in a stochastic game where the payoff reflects their ability to balance two competing behavioral tendencies:
one related to the environment and the other to social interactions.
Accordingly, the interaction source term in \cref{eq:general-structure} can be decomposed as
\begin{equation}
    {\cal J}_h = {\cal J}_h^G + {\cal J}_h^P,\notag
\end{equation}
where ${\cal J}_h^G$ denotes the contribution arising from purely pedestrian–environment (geometric) interactions, while ${\cal J}_h^P$ represents the contribution due to pedestrian–pedestrian interactions.

%-----------------------------------------------------------------------
\subsubsection{Dynamics induced by pedestrian--environment interactions}
%-----------------------------------------------------------------------
In an evacuation process, the geometry of the domain plays a crucial role in determining the trajectories adopted by pedestrians. The interaction between pedestrians and the environment is modeled as the combined effect of attraction toward the exit and repulsion from walls and other physical boundaries.

Pedestrians aim to reach the designated exit by following the shortest possible path. 	For a pedestrian located at $\mathbf{r}$, the distance to the exit is defined as  
\begin{equation}\label{d_E}
	d_E(\mathbf{r}) = \min_{\mathbf{r}^*\in E} \|\mathbf{r} - \mathbf{r}^* \|,
\end{equation}
where $\|\cdot\|$ denotes the Euclidean norm in $\mathbb{R}^2$. The corresponding unit direction vector from $\mathbf{r}$ to the exit is denoted by ${\vec{\nu}}(\mathbf{r})$ (see \cref{fig:1-domain}). 

If a pedestrian at position $\mathbf{\bar{r}}$ moves in a direction $\theta_\ell$ that does not lead toward the exit, a collision with a wall may occur at the point $\mathbf{\bar{r}}_W(\mathbf{\bar{r}},\theta_\ell)$, located at a distance $d_W(\mathbf{\bar{r}},\theta_\ell)$; see \cref{fig:1-domain}.
To prevent this, the pedestrian adjusts their path along the unit tangent vector ${\vec{\tau}}(\mathbf{\bar{r}}, \theta_\ell)$ to $\partial\Omega$ at $\mathbf{\bar{r}}_W$ which is chosen to lead them toward the exit.

Taking into account that environmental effects combine attraction toward the exit with the avoidance of walls, and that these tendencies become stronger as pedestrians approach the exit and the boundary, respectively, we define a geometrically preferred vector as
\begin{equation}
\label{eq:theta_G}
\vec{\omega}_G(\mathbf{r},\theta_\ell) = \left[ 1 - d_E(\mathbf{r}) \right]\vec{\nu}(\mathbf{r}) + \left[ 1 - d_W(\mathbf{r}, \theta_\ell) \right] \vec{\tau}(\mathbf{r}, \theta_\ell).
\end{equation}
The direction $\theta_G$ of this vector represents the \emph{geometrically} preferred direction.
%whose direction $\theta_G$ represents the geometrical preferred direction.

However, direction updates are restricted to adjacent directions within the discrete set $I_\theta$.  More precisely, a pedestrian moving in direction $\theta_\ell$ may only transition to $\theta_{\ell-1}$, $\theta_\ell$, or $\theta_{\ell+1}$, with probabilities that favor the direction closest to $\theta_G$.  These transition probabilities are encoded by the table of games $\mathcal{G}=\{\mathcal{G}_{\ell}(h)\}_{\ell,h=1,\dots,N_\theta}$, where $\mathcal{G}_\ell(h)$ denotes the probability that a pedestrian moving in direction $\theta_\ell$ switches to direction $\theta_h$ due to environmental effects. 

The source term accounting for purely pedestrian–environment interactions is formulated based on the transition probabilities from the current direction $\theta_\ell$ to $\theta_h$ as
\begin{equation}\label{eq:ped-env}
\mathcal{J}_h^G[f](t, \mathbf{r}) = \mu[\rho] \left( \sum_{\ell=1}^{N_{\theta}} \mathcal{G}_\ell(h)\, f_\ell(t, \mathbf{r}) - f_h(t, \mathbf{r}) \right),
\end{equation}
where $\mu$ represents the interaction frequency between pedestrians and the domain boundaries. This frequency is assumed to decrease with the local density, as lower densities allow pedestrians to more easily perceive the presence of walls and exits. Thus, we model this behavior by setting $\mu[\rho] = 1 - \rho$.

%-----------------------------------------------------------------------
\subsubsection{Dynamics induced by pedestrian--pedestrian interactions}
%----------------------------------------------------------------------- 
Beyond the influence of the physical geometry, the decision-making process of a pedestrian is also affected by interactions with other pedestrians. 
These social interactions reflect two main behavioral tendencies: congestion avoidance and alignment with the local flow.

On the one hand, pedestrians opt to move toward less congested areas in order to facilitate their motion. 
This behavior is modeled by the unit vector $\vec{\gamma}(\theta_\ell,\rho)$, which identifies the direction that minimizes the directional derivative of the density $\rho$ at the pedestrian’s current position $\mathbf{r}$.
Conversely, pedestrians may base its movement on social alignment. In particular, when interacting with another individual moving in direction $\theta_m$, a pedestrian may adopt that direction. This effect is represented by the unit vector $\vec{\sigma}_m = (\cos\theta_m, \sin\theta_m)$.

Combining these two tendencies, pedestrian-pedestrian interactions are represented by the vector
\begin{equation}\label{eq:theta_P}
\vec{\omega}_P(\theta_\ell,\theta_m,\rho) = \varepsilon \vec{\sigma}_m + (1 - \varepsilon)\vec{\gamma}(\theta_\ell,\rho), 
\end{equation}
whose direction $\theta_P$ defines the {\it interaction-based} preferred direction.
The parameter $\varepsilon\in[0,1]$ balances the relative importance of alignment and dispersion.

As in the case of pedestrian-environment interactions, direction updates are modeled as probabilistic transitions restricted to adjacent states. These transitions are described by the game table $\mathcal{P}=\{\mathcal{P}_{\ell m}(h)\}_{\ell,m,h=1,\dots,N_\theta}$, where $\mathcal{P}_{\ell m}(h)$ denotes the probability that a pedestrian initially moving in direction $\theta_\ell$ switches to direction $\theta_h$ after interacting with another pedestrian moving in direction $\theta_m$. The transition probabilities favor directions closer to $\theta_P$, given by \Cref{eq:theta_P}.

The source term accounting for pedestrian-pedestrian interactions is given by
\begin{equation}
\begin{split}
\mathcal{J}_h^P[f](t, \mathbf{r}) = \eta[\rho] \Biggl( \sum_{\ell,m=1}^{N_{\theta}} \mathcal{P}_{\ell m}(h)[\rho]\, f_\ell(t, \mathbf{r})\, f_m(t, \mathbf{r}) \\
- f_h(t, \mathbf{r})\, \rho(t, \mathbf{r}) \Biggr)
\end{split}
\label{eq:ped-ped}
\end{equation}
where $\eta$ denotes the interaction rate, i.e., the frequency of binary encounters per unit time. This rate is assumed to increase with the local density, reflecting the fact that interactions become more frequent in crowded areas; hence, we take $\eta[\rho] = \rho$.

\begin{remark} 
The parameter $\varepsilon \in [0,1]$ in \cref{eq:theta_P} controls the relative importance of the two competing effects, reinforcing one or the other depending on the scenario under consideration. A value of $\varepsilon = 0$ corresponds to purely density-driven behavior, in which only the tendency to move toward less congested areas is taken into account, whereas $\varepsilon = 1$ corresponds to purely alignment-driven behavior, in which only the tendency to follow the stream is considered.
As noted by Agnelli et al. (2014)~\cite{Agnelli14}, this parameter can be interpreted as a measure of the panic level. It plays a crucial role in the numerical experiments. Hereinafter, $\varepsilon$ is referred to as the panic parameter.
\end{remark}

%%--------------------------------------------------------------------------
%Data assimilation framework
\subsection{Ensemble-based data assimilation} \label{letkf}
%%--------------------------------------------------------------------------
The main aim of data assimilation is to estimate the state vector $\v x_k\in \mathbb R^{N_x}$ at time $t_k$ of a physical system, using an approximate numerical model $\mathcal M$ and given partial and noisy observations $\v y_k\in \mathbb R^{N_y}$. 
Since observations are available at discrete times $t_k$, we use the subscript $k$ to denote time-discrete variables.
We also introduce an observation operator $\mathcal{H}$, which maps the state space into the observation space.
Neither the numerical model nor the observation operator is perfect: the former involves simplifications and numerical approximations of the underlying physical system, while the latter introduces inherent uncertainties in the measurement process. Accordingly, both are assumed to be affected by additive errors, leading to the following hidden Markov model
\begin{align}
    \textbf{x}_{k+1}    &= \mathcal M(\textbf{x}_{k}) + \gv \eta_k         \label{eq:markov_model},\\
    \textbf{y}_k        &= \mathcal H (\textbf{x}_k) + \gv \kappa_k,    \label{eq:observations}
\end{align}
where $\gv \eta_k$, the model error, is assumed to be a realization of $\mathcal N(\v 0, \v Q)$ (a normal distribution with mean $\gv 0$ and covariance matrix $\mathcal{\v Q}$), and $\gv \kappa_k$, the observation error, is assumed to be a realization of $\mathcal N(\v 0, \v R)$.

In ensemble-based data assimilation \cite{carrassi2018data}, one assumes that the prior information of the system state is given by an ensemble of numerical model predictions $\v x_k^{\text{f},(j)}$, where the superscript `\text{f}' denotes the forecast and $j$ the ensemble member index. The solution for the estimation problem is the posterior probability density given the sequence of observations, $p(\v x_k|\v y_k,\,...\,,\v y_0)$, which is approximated via a Monte Carlo approach by an ensemble $\left\{\v x_k^{\text{a},(j)} \right\}_{j=1}^{N_\text{e}}$ of $N_\text{e}$ possible states, where the superscript `\text{a}' denotes the analysis or estimate. %\ma{Assuming the predictions and the observations are represented by Gaussian statistics}

In this work, we use the Local Ensemble Transform Kalman Filter (LETKF) \cite{hunt07}. In contrast to standard ensemble-based data assimilation \cite{burguers98}, which requires inverting an $N_x \times N_x$ forecast error covariance matrix, the LETKF performs the analysis in the ensemble subspace. In that subspace, the covariance matrices are orders of magnitude smaller, with size $N_{\text{e}} \times N_{\text{e}}$, where $N_{\text{e}} \ll N_y \le N_x$. However, because $N_{\text{e}} \ll N_x$, the sample covariance matrix is rank-deficient, introducing spurious long-range correlations due to sampling errors and leading to an underestimation of the analysis covariance. The LETKF addresses this issue through domain localization \cite{hunt07}.

At each time $t_k$, we perform a data assimilation step at every grid point $\mathbf{r}_i$, using only nearby observations. Let $\mathbf{y}_k$ denote the \emph{full observation vector} at time $t_k$, corresponding to observation locations $\mathbf{r}_\ell$, $\ell=1,\dots,N_y$. We assume observations in all grid points, so the grid size is $N_y$ and is defined for both the state vector and the obsevational fields.
For each grid point $\mathbf{r}_i$, we define a \emph{localized observation vector} $\mathbf{y}_{i,k}$, whose components $y_{i_{\ell},k}$ correspond only to those observation points $\mathbf{r}_\ell$ located within a distance $R_L$ from $\mathbf{r}_i$, i.e.,
\begin{equation*}
   \| \v r_i - \v r_{\ell} \|  < R_L,
   %\label{eq:loc_radius}
\end{equation*}
where $R_L$ denotes the localization radius.

From now on in this subsection, we drop the time index $k$ to avoid cluttering the equations, though all variables remain implicitly time-dependent.

For each grid point $\mathbf{r}_i$, we also define a \emph{localized state vector} $\v x_i$, which represents the restriction of the full model state to a local domain around $\mathbf{r}_i$. Each ensemble member is associated with a localized forecast state $\v x_i^{\text{f},(j)}$, for $j=1,\dots,N_\text{e}$.

 In the LETKF, we work with the forecast ensemble \textit{anomalies} (or perturbations) in the state space, defined as the matrix $\v X_i^\text{f}$ whose $j$-th column is
\begin{equation}
    \v X_i^{\text{f},(j)} = \frac{\v x_i^{\text{f},(j)} - \bar{\v x}_i^\text{f}}{\sqrt{N_\text{e}-1}},
    \label{eq:space_anomalies}
\end{equation}

\noindent where $\bar{\v x}_i^\text{f}$ is the ensemble mean of the localized forecast states given by
\begin{equation}\label{eq.xf.mean}
\bar{\v x}_i^\text{f} = \frac{1}{N_\text{e}} \sum_{j=1}^{N_\text{e}} \v x_i^{\text{f},(j)}.
\end{equation}

For each grid point $\mathbf{r}_i$, we introduce a localized observation operator $\mathcal{H}_i$, which maps a localized state vector $\v x_i$ to the corresponding localized observation space, making it directly comparable with the localized observation vector $\v y_i$ defined above. For each ensemble member we define
\begin{equation}
\v y_i^{\text{f},(j)} = \mathcal{H}_i\left(\v x_i^{\text{f},(j)}\right), \quad j=1,\dots,N_\text{e},
\end{equation}
which represents the projection of the localized forecast state onto the localized observation space. The corresponding forecast anomaly matrix $\v Y_i$ and mean forecast vector $\bar{\v y}_i^\text{f}$ in observation space are defined analogously to their state-space counterparts in \reff{eq:space_anomalies} and \reff{eq.xf.mean}, respectively.

We seek an analysis update similar to the original Kalman filter \cite{kalman60}, but performed in the ensemble space.

The analysis update for the ensemble mean in the ensemble space is given by
\begin{equation}
    \bar{\v x}_i^\text{a} = \bar{\v x}_i^\text{f} + \v X_i^\text{f}\v w_i^\text{a},
    \label{eq:kalman_gain}
\end{equation}
where the weight vector $\v w_i^\text{a}$ is computed as
\begin{equation}
    \v w_i^\text{a} = \v T_i^{-1} {\v Y_i}^\top \v R_i^{-1} \left( \v y^\text{f}_i - \bar{\v y}^\text{f}_i \right),
    \label{eq:wa}
\end{equation}
and 
\begin{equation}
    \v T_i = \left(\v I_{N_\text{e}} + {\v Y_i}^\top \v R_i^{-1}\v Y_i\right)^{-1/2},
    \label{eq:etmatrix}
\end{equation}
is the ensemble transform matrix with $\v I_{N_\text{e}}$ being the identity matrix of dimension $N_\text{e}\times N_\text{e}$ and $\v R_i$ the localized observation covariance matrix.

Finally, the full analysis ensemble is constructed by adding the updated anomalies to the updated mean
\begin{equation}
    \v x_i^{\text{a},(j)} = \bar{\v x}_i^\text{a} + \sqrt{N_\text{e}-1} \, \v X_i^\text{f} \v T_i^{(j)},
    \label{eq:posterior_etkf}
\end{equation}
where $\v T_i^{(j)}$ represents the $j$-th column of the ensemble transform matrix. 

While the analysis is performed locally at each grid point $i$, only the updated state at that specific location is retained from each local analysis -- the corrections to the surrounding grid points are discarded. This procedure is then repeated independently for every grid point in the spatial domain so that the global analysis state is assembled by collecting the local analysis at each grid point. The resulting mean global analysis, $\bar{\v x}^\text{a}$, provides our best point estimate of the system state.

A detailed derivation of the LETKF method from first principles may be found in Hunt et al. (2007) \cite{hunt07}.

%-------------------------------------------------------------------------------------------------------------------------------------------------------------------------
\subsection{State-parameter estimation}\label{subsec:augmented-approach}
%-------------------------------------------------------------------------------------------------------------------------------------------------------------------------
The state vector $\mathbf{x}_k$ contains all the information about the system at time $t_k$. In the context of the spatial kinetic model for crowd evacuation, the discrete model state, which we denote as $\v{f}_k$, consists of the distribution functions $f_h(t_k,\mathbf{r}_l)$ for $h = 1, \ldots, N_{\theta}$ and $l = 1, \ldots, N_{y}$.
Within the ensemble-based data assimilation framework, we adopt an \emph{augmented state} approach, in which the model parameters $\gv \psi$ are estimated jointly with the model state variables. To this end, we define the augmented state vector as $\v x_k = [\v f_k, \gv \psi_k]$.

Through this augmentation approach, the model parameters are estimated simultaneously with the state variables via the LETKF. In this framework, the parameters $\gv \psi$ act analogously to unobserved state variables: although they are not directly observed, they are corrected through the cross-correlations between the observed state variables and the parameters encoded in the augmented forecast covariance matrix. This joint estimation strategy in data assimilation is extensively discussed by Ruiz et al. (2013)\cite{Ruiz13}. The success of this approach depends on an accurate characterization of the augmented forecast covariance of the ensemble calculated in \cref{eq:posterior_etkf}. 

In this work, the parameter vector $\gv \psi$ reduces to a single scalar parameter $\varepsilon$, representing the global panic factor, although the methodology naturally uses local parameters and it is readily extended to use any number of parameters subject to its identifiability.

The LETKF framework assumes that all state variables are local. However, the parameter $\varepsilon$ is defined globally, which requires special treatment in order to be incorporated into the framework. To address this, we introduce a local representation by defining a parameter $\varepsilon_l$ at each grid point $\mathbf{r}_l$. These local parameters are then updated through the filter corrections at each valid grid point. The final global parameter estimate, $\varepsilon^a$, is defined as the spatial average of the local estimates, $\varepsilon^a = \sum_{l=1}^{N_y} \varepsilon_{l}^a / N_y$.

The local parameters $\varepsilon_{l}^a$ are updated only at grid points where the local density observation exceeds a prescribed threshold (set here to $0.05$). Grid points that do not satisfy this condition are excluded from the parameter update, as they are assumed to contain insufficient observational information. This restriction prevents spurious updates in low-density regions.

%-------------------------------------------------------------------------------------------------------------------------------------------------------------------------
\subsection{Obervations and observation operator}\label{obs_op}
%-------------------------------------------------------------------------------------------------------------------------------------------------------------------------
The increasing deployment of high-resolution video surveillance in public transit hubs and urban centers provides an unprecedented opportunity to apply data-driven methodologies to crowd flows. Data-driven frameworks allow a shift away from purely theoretical numerical simulations by using trajectory data to calibrate and validate complex evacuation models. Real-time crowd observations can be used to track individuals and reconstruct pedestrian trajectories frame by frame. 

In this study, we assume that the positions of pedestrians at each time frame are available from video surveillance, without identifying individuals or tracking their trajectories. The observations therefore consist of a discrete set of positions for each pedestrian $j$ at time $t_k$, denoted by $\{ r_x^{j}(t_k), r_y^{j}(t_k) \}_{j=1,\dots,N_0}$.

To transform these pedestrian trajectories into a continuous field, we apply kernel density estimation (KDE) using Gaussian kernels with a suitably chosen bandwidth. This procedure yields a smooth approximation of the crowd density from the raw position pedestrian data, which serves as an observation density field $\rho^{\text{o}}(t,\v r)$. The reconstructed density field is evaluated at the observation grid points $\v r_\ell$, $\ell=1,\dots,N_y$ to construct the observation vector 
$\v y_k=\left[\rho^\text{o}(t_k,\v r_\ell)\right]_{\ell=1,\dots,N_y}$.

The mapping from the augmented state vector to the observational space--defined by the observation operator--is given by the sum of the distribution functions, that is
\begin{equation}
\mathcal{H}(\mathbf{x}_k) = 
\begin{bmatrix} 
\sum_{h=1}^{N_\theta} f_h(t_k, \mathbf{r}_1) \\  
\vdots \\ 
\sum_{h=1}^{N_\theta} f_h(t_k, \mathbf{r}_{N_y}) 
\end{bmatrix}
\in \mathbb{R}^{N_y}.
\label{eq:obs_operator_rho}
\end{equation}

The observation operator defined above is linear. 

%-----------------------------------------------------------------------
\subsection{Inflation factor}
%
%-----------------------------------------------------------------------

Even with localization, the finite-size ensemble leads to sampling errors that cause the ensemble to systematically underestimate forecast uncertainty, so that the spread collapses over the cycles \cite{whitaker2002ensemble} and the filter may eventually diverge. To prevent this, we apply multiplicative inflation, in which the ensemble perturbations are scaled by an empirical inflation factor $\lambda$:
\begin{equation}
    \v x^{\text{a},(j)} \rightarrow \v x^{\text{a},(j)} + \lambda(\v x^{\text{a},(j)} - \bar{\v x}^\text{a}).
    \label{eq:mult_inf}
\end{equation}

Within the augmented state formulation, different inflation factors are applied to the state components $f_h(t_k,\v r_l)$ and to the parameters $\gv \psi$, motivated by their different dynamical behavior: while the former evolve according to the underlying dynamical system, the latter remain unchanged during the forecast step and are only updated through assimilation. Using a single inflation factor for all components would therefore cause a rapid collapse of the parameter ensemble spread and filter divergence \cite{Ruiz13b}.

%%------------------------------------------------------------------------------------------------------------------------------------------------------------
\section{Experimental details} \label{detalles_exp}
%%------------------------------------------------------------------------------------------------------------------------------------------------------------
 
\Cref{eq:general-structure} is solved numerically using an operator splitting method~\cite{holden2010}. Specifically, the governing equation is decomposed into a two-dimensional transport equation and a nonhomogeneous ordinary differential equation in time accounting for the source term.
The homogeneous transport equation is solved using the Python library \textit{PyMPDATA}~\cite{pympdata}, a high-performance implementation of the MPDATA algorithm originally introduced by Smolarkiewicz et al.~\cite{Smolarkiewicz98}. The non-homogeneous equation is solved using the explicit Euler method.

In all numerical experiments, we consider a square domain of side length $10 \, \mathrm{m}$, representing a room with a single exit door of width $1.5 \, \mathrm{m}$ located on the right boundary; see \cref{fig:1-domain}, and the velocity space is composed of $N_\theta = 8$ directions. 
A nondimensional formulation is adopted by scaling spatial coordinates with the domain diameter $L = 10\sqrt{2}$, while density and velocity are normalized with respect to $\rho_M = 7\,\mathrm{ped/m^2}$ and $V_M = 2\,\mathrm{m/s}$, respectively. The numerical solution of \cref{eq:general-structure} is obtained using a uniform spatial discretization with $\Delta r_x = \Delta r_y = 1/(100\sqrt{2})$, resulting in a discretized  state $\v f_k \in \mathbb{R}^{N_\theta \times N_y}$, where $N_\theta \times N_y = 80000$.
The numerical time step is set to $\Delta t = 0.01 \, \mathrm{s}$.

%%--------------------------------------------------------------------------
\subsection{Twin experiments} \label{twin_exp}
%%--------------------------------------------------------------------------

The first experiment conducted to evaluate the convergence and parameter identifiability of the DA framework consists of attempting to reproduce synthetic observations. These observations are generated by adding noise to the state given by the evolution of the nature model, \cref{eq:general-structure}, using fixed parameters. The prediction model used in the data assimilation process is identical to the nature model; however, it is initialized with different initial conditions for the density and the parameter $\varepsilon$. This implies that both the initial state and $\varepsilon$ are treated as unknown variables. In the context of data assimilation, this setup is formally known as 
twin experiments. These experiments are useful for evaluating whether the density and $\varepsilon$ estimates converge to the true values throughout the assimilation cycles. 
To this end, the ``nature'' evolution is defined using $\varepsilon = 0.35$ and a total population of $N_0 = 46$ pedestrians, divided into two groups. These two groups are initially located on opposite sides of the room and follow spatial Gaussian pattern, with initial velocities aligned with the directions $\theta_3$ and $\theta_7$, respectively (directed toward each other).
Setting the coordinate origin at the lower-left corner, the two groups are centered at $\mathbf{r}_1 = (3.0,2.0) \,\text{m}$ and $\mathbf{r}_2 = (2.0,6.5)\,  \text{m}$, both with a spatial standard deviation of $2.2 \, \text{m}$.

The ensemble is composed of $N_{\text{e}} = 40$ members, generated by introducing random perturbations to the mean and standard deviation of the true initial distribution.
The ensemble member mean positions of the two groups, denoted by ${\mathbf{r}}_1^{(j)}$ and $\mathbf{r}_2^{(j)}$, are sampled from the bivariate normal distributions $\mathcal{N}(\mathbf{r}_1, \sigma^2 \mathbf{I})$ and $\mathcal{N}(\mathbf{r}_2, \sigma^2 \mathbf{I})$, respectively, where $\sigma = 0.6\,\mathrm{m}$ and $\mathbf{I}$ denotes the $2 \times 2$ identity matrix. The spatial standard deviation of each group in each ensemble member is sampled from $\mathcal{N}\left(2.2\,\text{m}, 0.04\,\text{m}^2\right)$. The parameter $\varepsilon$ is drawn from $\mathcal N\left(0.6, 0.01\right)$.

We perform $300$ assimilation steps over $1800$ model time steps (one observation every $6$ forecasts), which corresponds to a total assimilated time length of $18\,\mathrm{s}$.

For the twin experiments, the true underlying dynamics of the studied system are known, so there is no need to use  the KDE observation treatment of \cref{obs_op}. Therefore, the observations are generated directly from the true density $\rho^\text{t}(t_k,\v r_l)$ at each observation time and grid point adding a multiplicative Gaussian noise:
\begin{equation}
\v{y}_{k,l} = \rho^o(t_k,\v r_l)= (1 +  \xi_{k,l}) \rho^\text{t}(t_k,\v r_l),
\label{eq:obs_noise} 
\end{equation}
where the noise $\xi_{k,l}$ is sampled from the distribution $\mathcal N(0,0.01)$. Unlike the additive noise formulation in \cref{eq:observations}, a multiplicative approach ensures that the noise amplitude is modulated by the local density. This prevents the signal from being dominated by noise in regions where the density is zero or near-zero, thereby maintaining a physically consistent observation error across the domain.

The inflation factors are set to $\lambda = 1.05$ for the density variables and $\lambda = 1.02$ for the panic parameter $\varepsilon$, while the localization radius is set to $R_L = 1~\text{m}$.  These values were obtained through a brute-force search, selecting the inflation factors and localization radius that yielded the best filter performance.

%%--------------------------------------------------------------------------
\subsection{ABM-generated observation experiments}\label{data_reales}
%%--------------------------------------------------------------------------

To evaluate the performance of the ensemble-based data assimilation framework in an environment where observations do not originate from the KTAP equations, we emulate pedestrian evacuation trajectories using an agent-based model (ABM) to generate synthetic observations. In this setting, the KTAP equations serve as a surrogate model, and data are assimilated in the presence of structural model errors.

The ABM simulates the spatial movement of individuals within the crowd, accounting for interactions both among pedestrians and with the surrounding environment. The position of each agent in the domain, denoted by $\v r^j$ for $j = 1, \ldots, N_0$, where $N_0$ is the initial number of agents, evolves according to the following system of  equations:
	
\begin{equation}
    \begin{aligned}
    {r_x}^{j}(t+\Delta t) &= {r_x}^{{j}}(t) +  v^{{j}} {\left[ \rho(t,\textbf{r}^{j}) \right]}g_x(\theta_G^{j}, \theta_P^{j}) \, \Delta t,
       \\ 
    {r_y}^{{j}}(t+\Delta t) &= {r_y}^{{j}}(t) +  v^{j} {\left[ \rho(t,\textbf{r}^{j}) \right]}g_x 
    (\theta_G^{j}, \theta_P^{j})\, \Delta t, 
  \end{aligned}
\end{equation}
with
\begin{equation}
  \begin{aligned}
    g_x(\theta_G^{j}, \theta_P^{j}) &= w \cos \theta_G^{j} + (1-w) \cos \theta_P^{j}, \\ 
    g_y(\theta_G^{j}, \theta_P^{j}) &= w \sin \theta_G^{j} + (1-w) \sin \theta_P^{j}.
  \end{aligned}
\end{equation}
Here $\theta_G^{j}$ and $\theta_P^{j}$ denote the  geometrical and interaction-based preferred directions of agent $j$, respectively. These directions are obtained analogously to \cref{eq:theta_G} and \cref{eq:theta_P} but use a discrete approximation and stochastic rules. 
Furthermore, $\Delta t$ is the time step and
$v^{j}$ is the velocity magnitude of pedestrian $j$, which  decays exponentially with the local density $\rho(t,\textbf{r})$. The parameter $w$ weights the balance between the interaction-based and  geometrical directions.   
The spatial room dimensions of the ABM are the same as the kinetic model, but due to the random nature of its dynamics, it achieves different macroscopic crowd behaviors. Similar  models have been implemented for instance  by Knopoff et. al (2025) \cite{knopoff2025individual}. 
The objective of this experiment is to estimate the state of the crowd and the panic parameter in a more realistic situation.
The observations are generated using the KDE method described in \cref{obs_op}. 
Once the macroscopic density field is obtained from agent trajectories, we add random noise according to \cref{eq:obs_noise}.
For this assimilation experiment with ABM observations, the number of pedestrians, the initial ensemble conditions, the observation  times and the localization radius are kept the same as in \cref{twin_exp}.
The inflation factors used in this experiment are $\lambda=1.07$ for all state variables, and $\lambda=1.015$ for the panic parameter $\varepsilon$, which were obtained via a brute-force search.

%------------------------------------------------------------------------------------------------------------------------------------------------------------
%Results
\section{Results} \label{resultados}
%%-----------------------------------------------------------------------------------------------------------------------------------------------------------

%%--------------------------------------------------------------------------
\subsection{Twin experiments} \label{exp_sinteticos}
%%--------------------------------------------------------------------------

\begin{figure}[H]
    \centering
    \includegraphics[width=0.65\textwidth]{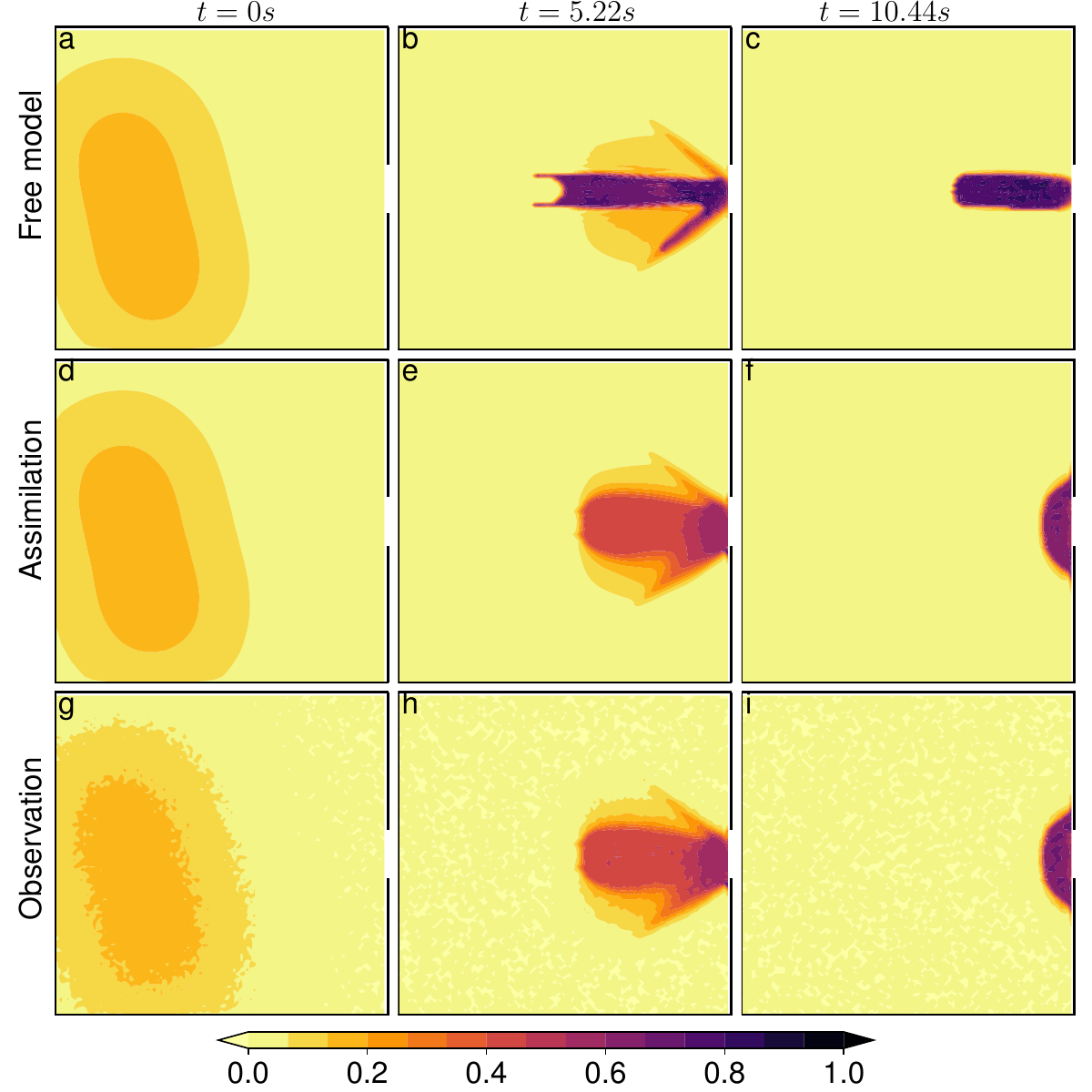}
    \caption{Evolution of the crowd density at different times ($t = 0$, $5.22$, and $10.40,\mathrm{s}$) for the free kinetic model with $\varepsilon = 0.6$ (a-c), the filter estimates (d-f), and the observations (g-i).}
    \label{fig:correcciones_twin}
\end{figure}

We first evaluate the framework's ability to simultaneously reconstruct the crowd state and identify the unknown parameter $\varepsilon$ in the twin experiments. \Cref{fig:correcciones_twin} compares the crowd density at different times for three cases: the free kinetic model initialized with $\varepsilon = 0.6$ (upper panels, a--c), the filter estimates (middle panels, d--f), and the synthetic observations (lower panels, g--i), generated from the nature model with $\varepsilon = 0.35$.

Large values of $\varepsilon$ produce a highly organized crowd, characterized by a strong tendency to ``follow the flock'' dynamics, whereas smaller values lead to more unorganized dynamics, with pedestrians spreading out and utilizing the full width of the exit. Despite the significant dynamical differences between the free model ($\varepsilon = 0.6$) and the observations ($\varepsilon = 0.35$), the filter successfully steers the model state to accurately reproduce the observations providing smoother fields than the noisy observations (\cref{fig:correcciones_twin}d--f).

\begin{figure}[H]
    \centering
    \includegraphics[width=.48\textwidth]{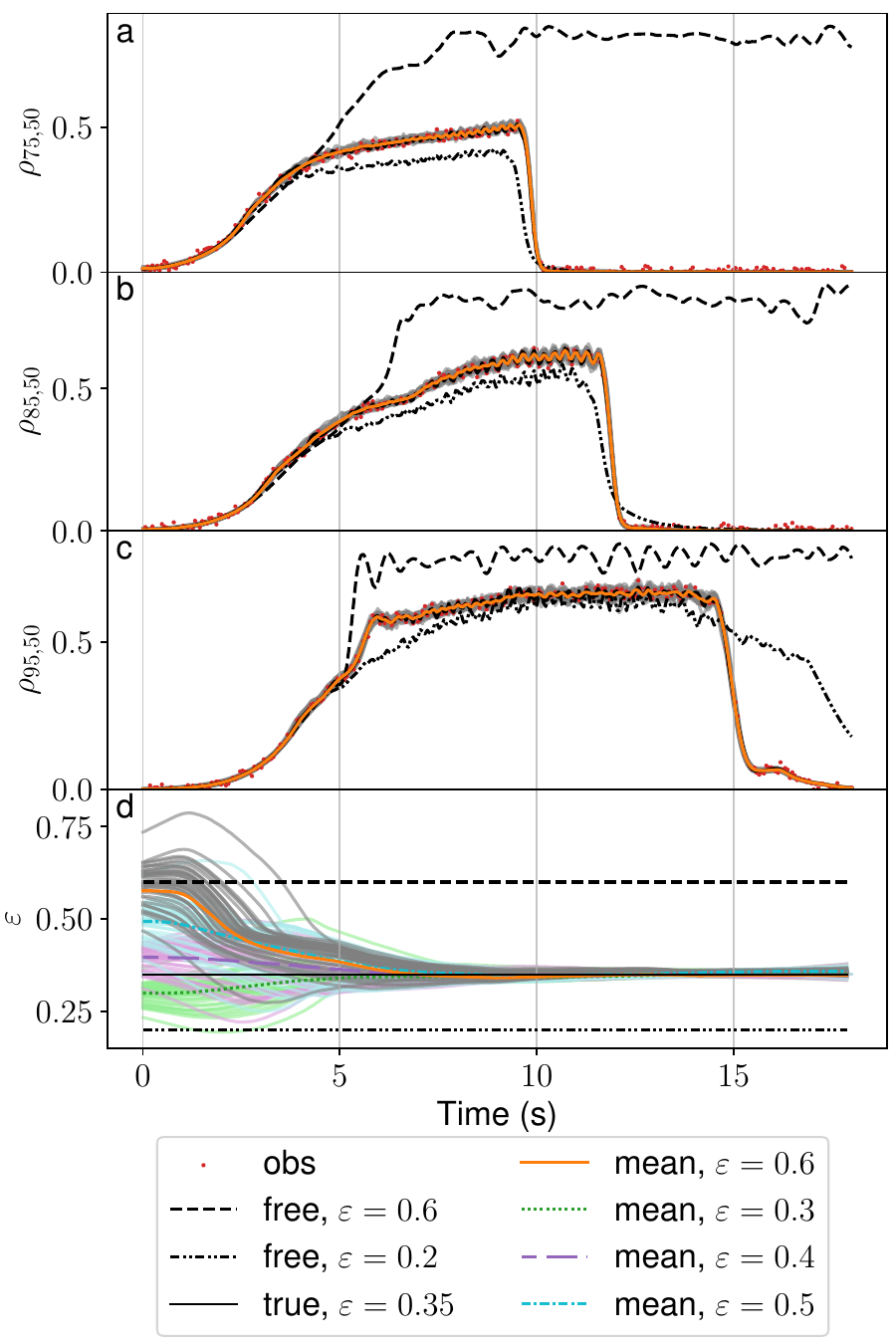}
    \caption{Evolution of the pedestrian density at three selected grid points ($\rho_{75,50}$, $\rho_{85,50}$, and $\rho_{95,50}$) as a function of time during the twin experiment (a-c). The gray curves denote the ensemble members, the orange curve their mean, the black solid curve the reference (true) solution, and the red dots the observations. The black dashed and dash-dotdotted curves correspond to free model simulations with fixed panic values $\varepsilon = 0.6$ and $\varepsilon = 0.2$, respectively.
    The temporal evolution of the estimated panic parameter $\varepsilon$ is presented in (d). The black solid curve indicates the true value, while the additional curves (green dotted, purple long-dashed, and cyan dash-dotted) correspond to estimates obtained from different initializations of $\varepsilon$.}
    \label{fig:series_temporales_twin}
\end{figure}

\Cref{fig:series_temporales_twin}(a--c) show the temporal evolution of the pedestrian density $\rho(\v r,t)$ at three selected grid points ($(75,50)$, $(85,50)$, and $(95,50)$) near the center of the exit, which we denote by $\rho_{75,50}$, $\rho_{85,50}$, and $\rho_{95,50}$.
\Cref{fig:series_temporales_twin}(d)  displays the corresponding estimates of the panic parameter as a function of time. 
In all subfigures, the gray curves represent the ensemble members, the orange curve denotes the ensemble mean, the black solid curve represents the reference (true) solution, and the red dots indicate the observations. The black dashed and black dash-dot-dotted curves correspond to integrations of a free model initialized with the same initial pedestrian distribution of the true model, but with fixed values of $\varepsilon = 0.6$ and $\varepsilon = 0.2$, respectively.

The density evolution shown in panels (a--c) firstly reveals that the two free simulations diverge significantly from the reference solution, highlighting the strong influence of the panic parameter on pedestrian dynamics: different values of $\varepsilon$ lead to markedly distinct crowd behaviors. Besides, the results demonstrate that the data assimilation framework accurately reconstructs the density evolution, with the filter effectively tracking the observations and providing accurate estimates of the state. \Cref{fig:series_temporales_twin}(d) shows that the estimate of the parameter $\varepsilon$ converges to the true value ($\varepsilon = 0.35$) at approximately $t = 6\,\text{s}$. In addition, we conducted independent assimilation experiments using different initial guesses for the panic parameter (green dotted, purple long-dashed, and cyan dash-dotted curves), all of which succesfully converged to the true value. This demonstrates that the panic parameter is identifiable within the ensemble-based data assimilation framework and that its estimation is robust with respect to the initial guess. Furthermore, as more observations are assimilated, the parameter not only converges, but its uncertainty---quantified by the ensemble spread---also decreases over time.

\begin{figure}[H]
    \centering
    \includegraphics[width=.65\textwidth]{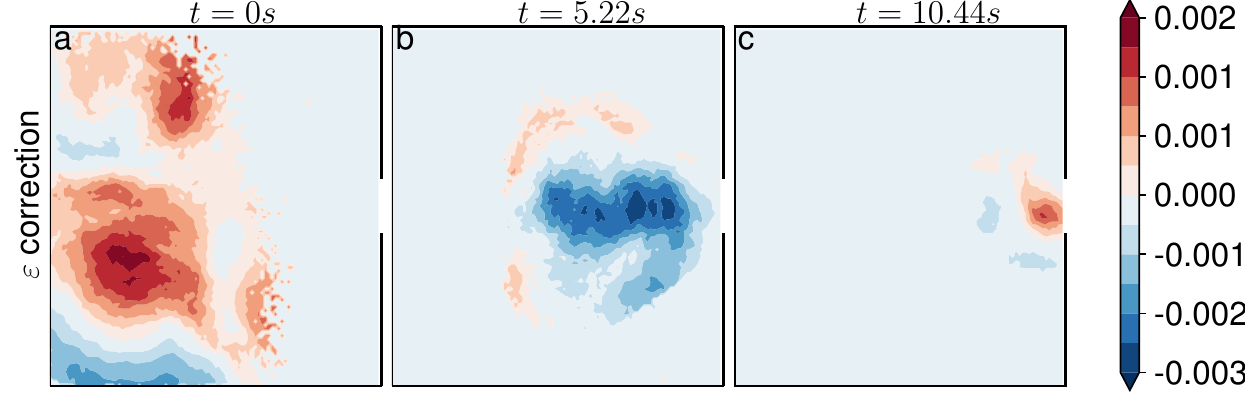}
    \caption{Spatial correction $\varepsilon^\text{a}_{k,l}-\varepsilon^\text{f}_{k,l}$ of the panic parameter $\varepsilon$  given the crowd observations. The filter is initialized with a mean value $\varepsilon=0.6$.
    }
    \label{fig:snapshots_twin}
\end{figure}

\Cref{fig:snapshots_twin} (a--c) show the spatial correction that the filter performs on $\varepsilon$ at different simulation times, $\varepsilon_{k,l}^a-\varepsilon_{k,l}^f$. It is observed that corrections to the parameter occur with greater intensity at points where density is high. As the simulation progresses, the magnitude of the correction decreases as the parameter converges to the true value. Negative values in \cref{fig:snapshots_twin} (a--c) indicate that the estimated value of $\varepsilon$ decreases, and positive values indicate that the estimated parameter increases with respect to the forecast value. The net global parameter corrections are positive at $t = 0$, but overall negative at intermediate times. This effect in DA is known as \textit{spin-up}, which is an adjustment process caused by the difference between obsevations and the forecast model \cite{Kalnay03}. The spin-up time can be seen in \cref{fig:series_temporales_twin} (d), where the parameter correction starts after a few seconds before stabilizing.

\begin{figure}[H]
    \centering
    \includegraphics[width=.65\textwidth]{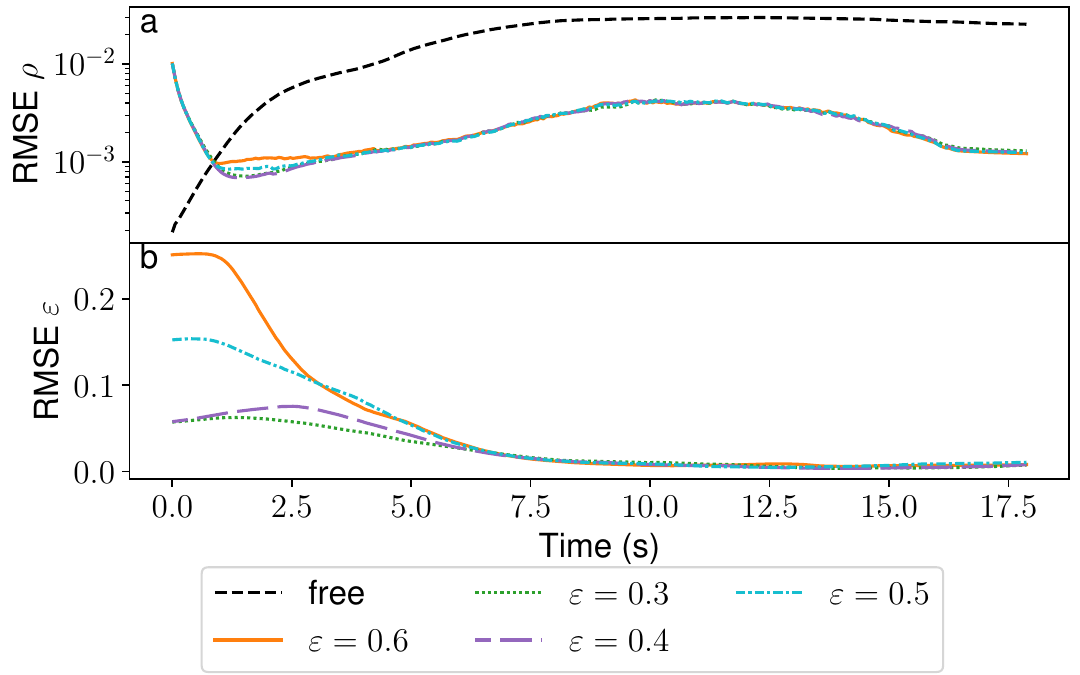}
    \caption{Root mean square error (RMSE) of the estimated density (a) and the panic parameter $\varepsilon$ (b) compared to the true state, as a function of time in the twin experiment. Different colors and line styles correspond to assimilation experiments with different fixed values of $\varepsilon$. The black dashed curve in panel (a) represents the RMSE of the free simulation with $\varepsilon = 0.6$.}
    \label{fig:rmse_twin}
\end{figure}

\Cref{fig:rmse_twin}(a) compares the temporal evolution of the RMSE for the density estimation under different initializations of the parameter $\varepsilon$ (green dotted, purple long-dashed, and cyan dash-dotted curves), with respect to the true state. All assimilation experiments using different initial guess values of $\varepsilon$ exhibit similar overall errors. For reference, the black dashed curve represents the RMSE of the free simulation with a fixed panic parameter of $\varepsilon=0.6$, respectively. Since this simulation is initialized with the true initial pedestrian distribution, its RMSE is zero initially; however, as time progresses, the free simulation diverges from the true state, whereas all the DA experiments give smaller  RMSEs  and similar between them independently of the initial parameter. \Cref{fig:rmse_twin}(b) shows the RMSE of the panic parameter estimation under different initial guess parameter values. The green dotted, purple long-dashed, cyan dot-dashed and orange curves correspond to $\varepsilon=0.3$, $\varepsilon=0.4$, $\varepsilon=0.5$ and $\varepsilon=0.6$ initializations, respectively. In all cases, the RMSE decreases over time as the parameter converges toward the true value, consistent with the results shown in \cref{fig:series_temporales_twin}(d). 

%%--------------------------------------------------------------------------
\subsection{Experiments with ABM-generated synthetic observations} \label{exp_reales}
%%--------------------------------------------------------------------------

\begin{figure}[H]
    \centering  
    \includegraphics[width=.65\textwidth]{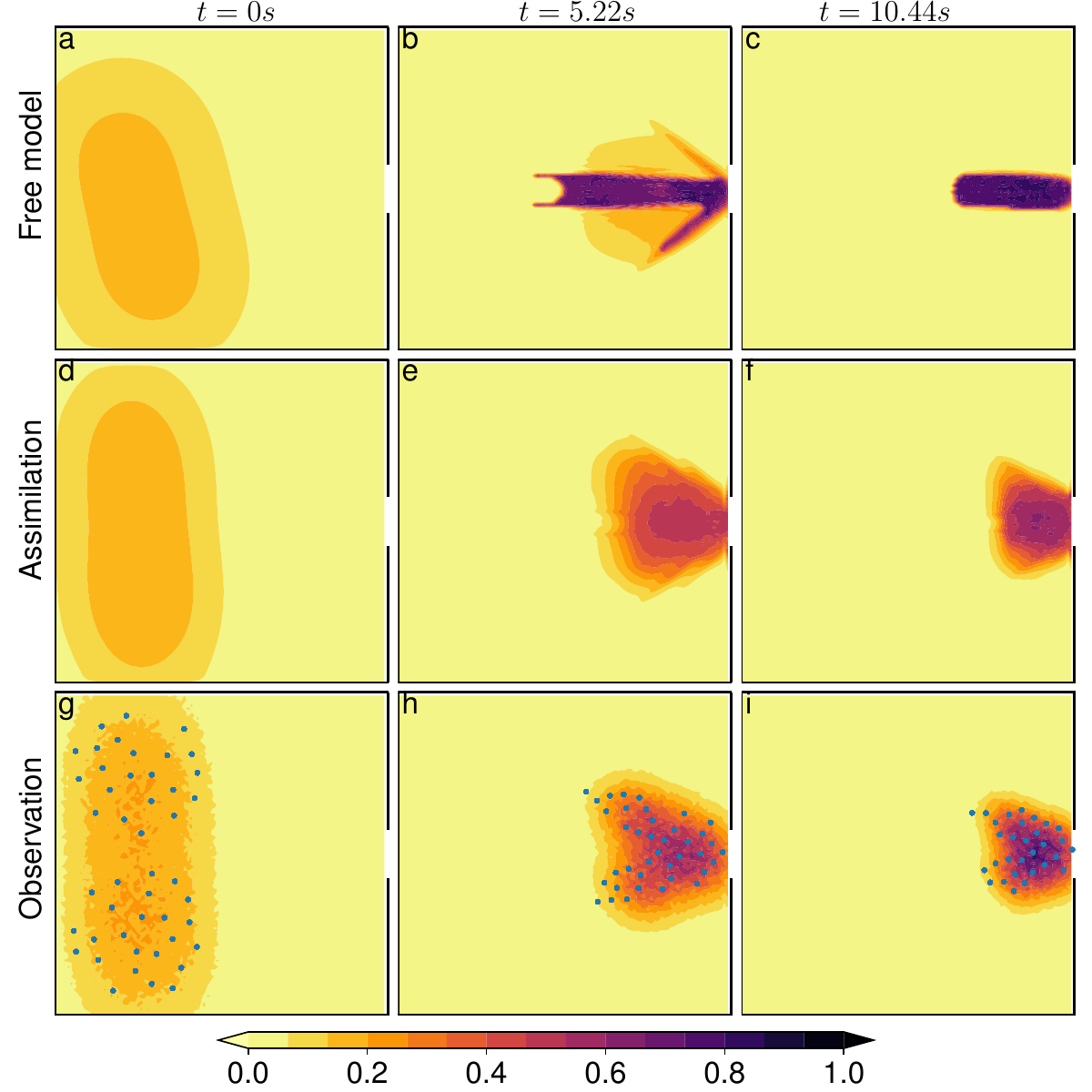}
   \caption{Density evolution is shown at times $t = 0$, $t = 5.22$, and $t = 10.44$~s for the free mesoscopic simulation with a fixed parameter $\varepsilon = 0.6$ (a--c);  the density estimated by the DA filter (d--f); and the ABM pedestrian locations  along with the corresponding derived observation fields (g--i).}
    \label{fig:correcciones_abm}
\end{figure}

\Cref{fig:correcciones_abm} (a--c) compare the evolution of the density at times $t=0$, $t=5.22$ and $t=10.44$~s of the free kinetic model initialized with $\varepsilon=0.6$, the filter correction (d--f), and the ABM observations (g--i). Examining the differences between the reference simulations in \Cref{fig:correcciones_twin}(g--i) and \Cref{fig:correcciones_abm}(g--i), we observe that the mesoscopic model tends to accumulate density over a wider area but closer to the exit. In contrast, the ABM density pattern remains more extended further from the exit and is directly aligned with the physical width of the exit. Despite these underlying dynamical differences, the filter successfully steer the model state to accurately reproduce the observations \cref{fig:correcciones_abm} d--f).

\begin{figure}[H]
    \centering
    \includegraphics[width=.48\textwidth]{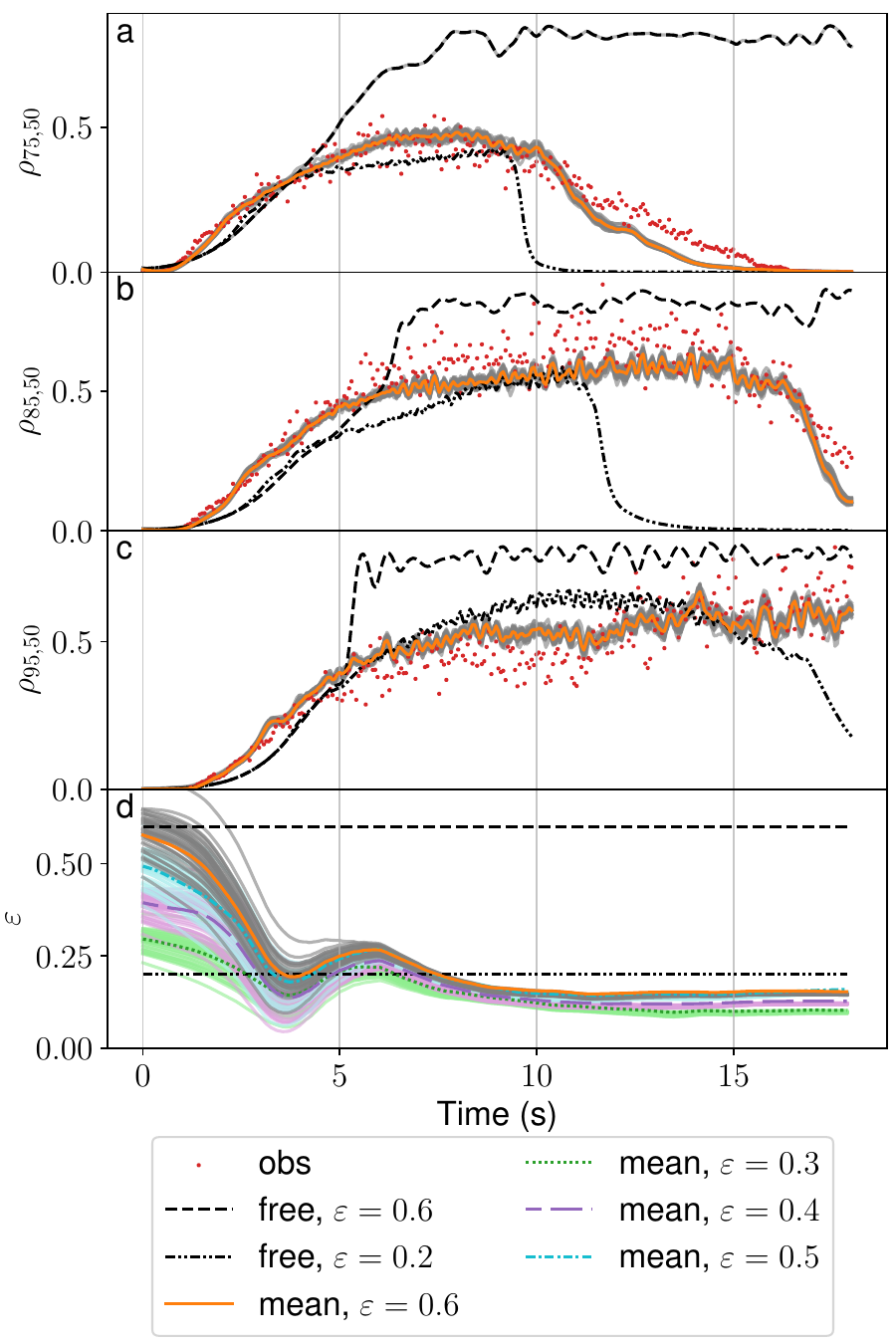}
    \caption{Evolution of the pedestrian density at three selected grid points ($\rho_{75,50}$, $\rho_{75,50}$ and $\rho_{95,50}$) as a function of time for the ABM observation experiment (a-c). Gray curves represent the ensemble members, the orange curve indicates their mean, red dots denote the observations and the black dashed and black dash-dot-dotted curves are reference free runs with a fixed panic value of $\varepsilon=0.6$ and $\varepsilon=0.2$, respectively. The temporal estimation of the parameter $\varepsilon$ is  presented in (d), where  the additional curves (green dotted, purple long-dashed, and cyan dash-dotted) represent the estimates obtained from alternative initializations of $\varepsilon$.}
    \label{fig:series_temporales_abm}
\end{figure}

\Cref{fig:series_temporales_abm}(a--c) show the temporal evolution of the pedestrian density $\rho(\v r,t)$ at the three selected grid points ($\rho_{75,50}$, $\rho_{85,50}$, and $\rho_{95,50}$), while panel (d) displays the estimated parameter $\varepsilon$. The curve styles and color conventions follows those defined for the twin experiment. As expected, the discrepancies between the two free mesoscopic simulations and the ABM observations are even more pronounced in this experiment compared to the twin experiment, owing to the presence of structural model error. In general, the filter  accurately tracks the observations; but discrepancies emerge at some locations and long times as seen in \cref{fig:series_temporales_abm}(a)). This is primarily a consequence of the intrinsic differences in the underlying dynamics, the mesoscopic model concentrating the density pattern closer to the exit at long times $t>11 s$. The filter is not able to fully correct these differences. \Cref{fig:series_temporales_abm}(d) shows that the parameter $\varepsilon$ converges toward a low, nearly constant value across different initializations in the ABM experiment. This suggests that smaller values of $\varepsilon$ allow the kinetic model to better reproduce the density patterns generated by the ABM. This effectively optimized parameter value may be compensating for other unresolved physical discrepancies between the ABM and the mesoscopic framework, such as a distinct tendency in the latter to alter alignment dynamics in low-parameter regimes. Parameter corrections may frequently compensate for structural differences between a surrogate model and the true nature system~\cite{Ruiz15}.

\begin{figure}[H]
    \centering
    \includegraphics[width=.65\textwidth]{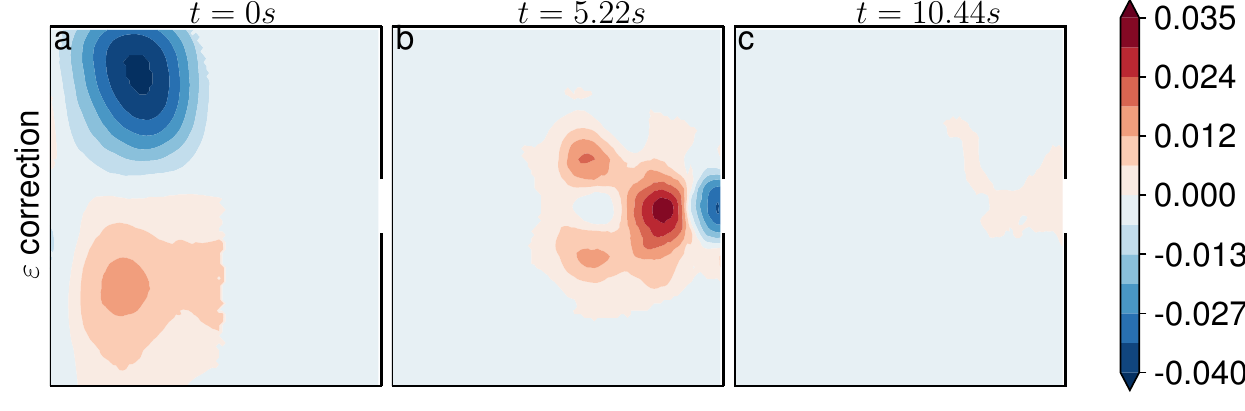}
    \caption{Spatial correction $\varepsilon^\text{a}_{k,l}-\varepsilon^\text{f}_{k,l}$ of the panic parameter $\varepsilon$ given the crowd observations. The filter is initialized  with a mean value of $\varepsilon=0.6$, in the ABM obsevations experiment.} 
    \label{fig:snapshots_abm}
\end{figure}

\Cref{fig:snapshots_abm} (a--c) show the spatial analysis correction performed by  the filter on the parameter $\varepsilon$ at three fixed times.
The net global parameter corrections are negative at $t = 0$. At intermediate times, however, the estimated parameter shows a slight increase, with corresponding net positive corrections at $t = 5.22$, followed by a slight net decrease.
This behavior matches the parameter estimation observations of Rosa et al. (2025) \cite{Rosa25}, where a set of critical parameters for a comportmental epidemiological model was estimated using an Ensemble Kalman Filter (EnKF). In that study, the authors noted that certain parameters exhibited a temporal tendency  reminiscent of a damped oscillation—undergoing transient fluctuations and corrections—prior to achieving convergence.

\begin{figure}[H]
    \centering
    \includegraphics[width=.65\textwidth]{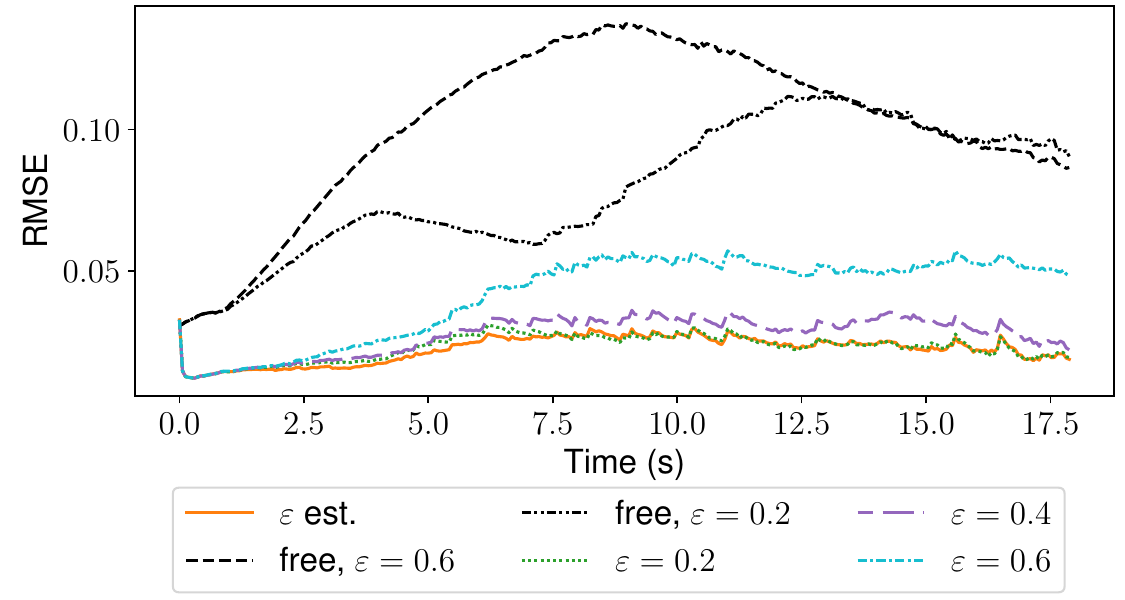}
    \caption{RMSE of the estimated density with respect to the observations as a function of time in the ABM-generated observation experiment. The orange curve corresponds to the case in which the parameter $\varepsilon$ is dynamically estimated, while the cyan dash-dotted, purple dashed and green dotted curves represent assimilation experiments with fixed values $\varepsilon = 0.6$, $0.4$, and $0.2$, respectively. The black dashed and black dash-dot-dotted curves show the RMSE of the free kinetic model with $\varepsilon=0.6$ and $\varepsilon=0.2$, respectively.}
    \label{fig:rmse_abm}
\end{figure}

We also conducted a series of sensitivity experiments to evaluate the impact of parameter estimation on filter performance. In these experiments, the density field is estimated within the filter while the panic parameter $\varepsilon$ remains fixed, rather than being dynamically updated within the augmented state space. \Cref{fig:rmse_abm} shows the temporal evolution of the $\text{RMSE}$ for the density estimation across these experiments. The case where $\varepsilon$ is dynamically estimated (orange curve) is contrasted against scenarios where $\varepsilon$ is held fixed at $0.6$ (cyan dashed-dotted curve), $0.4$ (purple dashed curve), and $0.2$ (green dotted curve). Additionally, the black dashed and black dash-dot-dotted curves provide baselines, representing the $\text{RMSE}$ of a free run of the kinetic model with a fixed panic value of $\varepsilon = 0.6$ and $\varepsilon = 0.2$, respectively, without data assimilation. The assimilation with parameter estimation gives the minimum RMSE with respect to the fixed parameter assimilations. The experiment with high panic parameter $\varepsilon=0.6$ (cyan curve) significantly compromises the overall density estimation, while the experiment with a parameter close to the optimal one ($\varepsilon=0.2$, green curve) gives comparable RMSE for long times. Crucially, while  calibration techniques are fundamentally limited to identifying a single, static parameter value for the entire simulation window, the augmented-state estimation framework successfully tracks a time-varying parameter. This capability allows the system to capture transient behavioral shifts in the crowd dynamics that a fixed, pre-calibrated value cannot represent.

\Cref{fig:rmse_abm} also shows the impact of assimilating observations into a complex system using an approximated model. The free simulation with a fixed $\varepsilon=0.6$ is shown by the dashed black curve, while  the cyan curve corresponds to the assimilation experiment using the same fixed parameter value. As can be seen, the free simulation exhibits a significantly larger RMSE (much larger than in the twin experiment, see \cref{fig:rmse_twin}), which is a direct product of structural model error (and nonoptimal parameters). A similar analysis holds for the comparison between the free simulation with $\varepsilon=0.2$ (black dash-dot-dotted curve) and the assimilation experiment with the same fixed parameter (green dotted curve) or the estimated parameter (orange curve).
While these dynamical discrepancies highlight the inherent limitations of the surrogate model, the data assimilation framework successfully drives the system state toward the observations. By correcting the forecast states based on observation-driven innovation terms, the filter effectively synchronizes the approximated model evolution with the observed pedestrian dynamics. This ability to compensate for model deficiencies and maintain a trajectory close to the natural evolution is a key advantage when estimating complex interaction parameters.

%%------------------------------------------------------------------------------------------------------------------------------------------------------------
%Conclusions
\section{Conclusions} \label{conclusiones}
%%------------------------------------------------------------------------------------------------------------------------------------------------------------

In the kinetic theory of active particles, the system is described by transport-type partial differential equations in which microscopic interactions are accounted for through suitable source terms. In this work, we propose a data-driven framework to model KTAP systems based on an ensemble-based data assimilation technique. Within this approach, the governing equations are sequentially coupled to the observational data as the system evolves; namely the state is corrected at each observation time using a combination of model and observation error statistics. This recursive process results in an estimated state trajectory that is constrained to the ground truth dynamical manifold. Because the underlying model is governed by spatio-temporal partial differential equations, ensemble-based data assimilation provides a natural and effective framework for integrating model predictions with observational data. The proposed methodology is presented and assessed through two representative numerical experiments.

The proposed framework is designed to jointly estimate the density field and a global panic parameter within the kinetic crowd model. The parameter estimation was achieved using a novel augmented state technique that relies exclusively on local observations. Through twin experiments, we demonstrated the identifiability of the panic parameter and the stable convergence of the method to the true value. Crucially, the estimation of this panic parameter is also robust to different initial guess parameter values, all experiments converging to the true value  within the same time scale. The results show that the DA system effectively steers the mesoscopic model toward the observed states. 

To further validate the approach, we assimilated synthetic observations generated by an ABM that simulates a realistic room evacuation through individual decision-making agents. The DA framework successfully constrained the kinetic model to track the ABM observations, effectively overcoming the inherent dynamical differences between the microscopic and mesoscopic scales of the two involved models. Ultimately, the joint estimation of the density field and the panic parameter achieved stable convergence to the ground truth trajectory and substantially reduced the overall estimation error compared to free KTAP simulations, leading to dynamics that differ significantly from those of a free-running model. This demonstrates that the DA framework can effectively bridge the gap between individual-level stochastic behaviors and the continuous density representations of the kinetic theory. Furthermore, the successful recovery of the panic parameter suggests that the method is capable of identifying latent behavioral parameters that are otherwise difficult to measure directly from survelliance data.

Both the inverse problem approach developed by Kim et al. (2025)\cite{Kim25} and the proposed ensemble-based data assimilation aim to account for a data-driven formulation within KTAP framework. The methodology in Kim et al. (2025)\cite{Kim25} treats parameter estimation as a constrained optimization problem, where the kinetic model acts as a formal constraint to minimize the discrepancy between simulated and observed density fields. In their approach, the panic parameter is optimized across both space and time. In contrast, our ensemble-based technique utilizes an augmented state that incorporates both the system state and the model parameters. This framework explicitly accounts for the uncertainty in the observations, model predictions and parameters, with the latter two being dynamically quantified via the emperical covariance of the ensemble representation. This makes the ensemble framework particularly effective for validating kinetic models against noisy surveillance data, as it can effectively steer the surrogate mesoscopic kinetic representation to track microscopic trajectories despite the inherent dynamical differences arising from model simplifications.

A promising direction for future work involves the objective estimation of the functional forms within the interaction operators. Pulido et al \cite{Pulido16} presented a methodology based on ensemble-based data assimilation to determine the functional forms of source terms in the system equations. This methodology  offers a means to infer the underlying physics of pedestrian movement directly from data. Applying this methodology would enable, for instance, the direct estimation of the functional dependence of pedestrian speed $v[\rho]$ as a function of density using surveillance data. Such an extension would transition the model from using predefined behavioral rules to a more flexible, data-driven representation of the geometric-pedestrian and pedestrian-pedestrian interaction terms of the source operator, $\mathcal{J}_h [f]$.

\section*{Acknowledgments}
The authors acknowledge the High-Performance Computing Center (CECONEA) at UNNE for providing the computational resources that supported this work.

\bibliographystyle{unsrt}
\bibliography{references}

\end{document}